\documentclass[twocolumn,amsmath,amssymb,showpacs,prb]{revtex4-1}
\usepackage{bm}
\usepackage{graphicx}

\usepackage{color}

\begin{document}
%\preprint{arXive:11}
%{\bf draft1\_5.tex}\\

\title{Enhancement of Antiferromagnetic Correlations below Superconducting Transition Temperature in Bilayer Superconductors}
%\title{Induce Effects of Antiferromagnetic Correlations 
%in Bilayer Superconductors}
\author{Hiroyuki Yoshizumi}\email{ydsumi@yukawa.kyoto-u.ac.jp}
\affiliation{
Yukawa Institute for Theoretical Physics, Kyoto University, 
Kyoto 606-8502, Japan}

\author{Takao Morinari}
\altaffiliation[Present address: ]{Graduate School of Human and Environmental Studies, Kyoto University, Kyoto 606-8501, Japan}
\affiliation{
Yukawa Institute for Theoretical Physics, Kyoto University, 
Kyoto 606-8502, Japan}

\author{Takami Tohyama}
\affiliation{
Yukawa Institute for Theoretical Physics, Kyoto University, 
Kyoto 606-8502, Japan}

\date{\today}

\begin{abstract}
  Motivated by the recent experiment in multilayered cuprate superconductors 
  reporting the enhancement of antiferromagnetic order 
  below the superconducting transition temperature, 
  we study the proximity effect of the antiferromagnetic correlation
  in a bilayer system and also examine the possibility
  of a coexistence of antiferromagnetic order and superconductivity.
  We present the result of mean field theory that is consistent
  with the experiment and supports the proximity effect picture.
\end{abstract}


\maketitle

\section{Introduction}
% high-Tc occurs near AFLRO
In the study of high-temperature superconductivity in the cuprates,
there has been intense interest in the interplay between
antiferromagnetism and superconductivity.
The undoped parent compound, which is a Mott insulator,
is an antiferromagnetic (AF) long-range ordered state.
Upon carrier doping, the AF state is converted into
the high-temperature superconducting (SC) state.
In the mechanism of superconductivity,
the AF correlation is believed to play an important role.
% multi-layered system -> Tc depends on the number of layers
In order to investigate the AF correlation effect
on superconductivity, multilayer systems are useful.
In multilayer systems, the number of CuO$_2$ layers
per unit cell, $n$, takes $n\ge 2$.
From systematic studies of the $n$ dependence of 
SC transition temperature $T_{\rm c}$,
it is found that the maximum $T_{\rm c}$ is obtained \cite{Ihara88,Parkin88}
for the case of $n=3,4$.
Since coupling between the CuO$_2$ planes
within the unit cell is stronger
than that between the CuO$_2$ planes in different unit cells,
the AF correlation can be strong in multilayer systems.
There is also charge imbalance over the CuO$_2$ layers
as revealed by the $^{63}$Cu Knight shifts, 
and thus the multilayer systems form a natural heterostructure 
at an intermediate carrier-concentration region.\cite{Tokunaga99}
%This imbalance is probably caused by large inter-layer electron hopping, 
%which make it possible to transfer doped carriers 
%from the apical site of the outer plane to the inner plane. 
%

% coexistence? NMR study
From nuclear magnetic resonance (NMR) experiments
%Recently NMR studies of multilayered cuprates
it was suggested that AF moments survived
even in a SC 
phase.\cite{Kotegawa04,Mukuda06,Mukuda08,Shimizu09}
A possibility of a phase separation is ruled out because 
the NMR signal associated with the paramagnetic state
was not observed.\cite{Mukuda06}
An important issue here is whether SC order
coexists with AF order~\cite{Inui88,Chen90,Giamarchi91} or not.
Recently, Shimizu {\it et al}. studied 
the temperature dependence of the AF moment in a five-layered cuprate Ba$_2$Ca$_4$Cu$_5$O$_{10}$(F,O)$_2$
by NMR measurements.\cite{Shimizu11}
They reported that the AF moment in the outer layer is enhanced below $T_{\rm c}$.
Such an enhancement of the AF moment below $T_{\rm c}$ does not appear in the coexistence phase of other superconductors such as iron arsenide superconductor Ba$_2$(Fe$_{1-x}$Co$_x$)$_2$As$_2$, where the AF moment reduces at $T_{\rm c}$.\cite{Fernandes10} Therefore, the enhancement seen in the five-layered cuprate is not necessarily common to all of the coexistence phase of superconductors. Noting that the iron arsenide is a single-layer system, we may speculate that the multilayer nature is essential for the enhancement, and thus need to clarify the mechanism of the AF moment enhancement triggered by the SC
transition under the presence of multilayers.

% AF moment enhancement by SC transition
In this paper, we study the interplay between SC and AF orders
within a mean field theory.
To describe a multilayer system,
where there are two types of layers associated with
outer planes and inner planes,
we consider a bilayer system 
with different electronic correlations in each layer.
By controlling interaction parameters 
for the AF correlation and the SC correlation,
we study the proximity effect between two layers with different order
and also examine the possibility of coexistence of SC and AF orders.
We find that the proximity effect leads to an enhancement
of AF order below the SC transition temperature 
that is consistent with the experiment.\cite{Shimizu11}
By contrast, we find qualitatively different behaviors
of the order parameters for the coexistence case.

This paper is organized as follows. 
In Sec.\ref{sec_model}, 
we introduce the bilayer Hamiltonian and describe the mean field theory.
In our model, a coexistence phase of SC and AF orders
is possible within a single layer.
This coexistence phase is described in Sec.\ref{sec_single}.
In Sec.\ref{sec_bilayer}, we describe the results 
about the proximity effect in the bilayer system.
We show that the temperature dependence of the AF moment
is consistent with the experiment.\cite{Shimizu11}
We also examine coexistence phases and show
that the temperature dependence of the order parameters 
is quite different from the experiment.
Finally, Sec.\ref{sec_summary} is devoted to summary and discussions.

\section{Model and Formalism}
\label{sec_model}
We consider a bilayer system with interactions 
which stabilize SC and/or AF order.
The two layers are coupled through an interlayer tunneling.
The Hamiltonian is given by
\begin{equation}
{\cal H} = \sum_{l=1,2} {\cal H}_l + {\cal H}_\perp,
\end{equation}
where the interlayer tunneling term ${\cal H}_\perp$ is
\begin{equation}
{\cal H}_\perp =
-t_p \sum_{\bm{k}, \sigma}
\left( c_{1,\bm{k} \sigma}^\dagger 
c_{2,\bm{k} \sigma} + \text{H.c.} \right).
\end{equation}
Here $c_{l,\bm{k} \sigma}^\dagger$ 
($c_{l,\bm{k} \sigma}$) 
creates (annihilates) electrons with 
in-plane momentum $\bm{k}$ and spin $\sigma$ at layer $l$.
We assume that the interlayer hopping matrix $t_p$ is independent of $\bm{k}$.
The Hamiltonian for the $l$ layer is
\begin{eqnarray} 
{\cal H}_l &=& 
\sum_{\bm{k}, \sigma} 
\xi_{\bm{k}} c_{l, \bm{k} \sigma}^\dagger c_{l, \bm{k} \sigma}
- V_l \sum_{j}
\left( 
c_{l j \uparrow}^\dagger c_{l j \uparrow} 
-c_{l j \downarrow}^\dagger c_{l j \downarrow} 
\right)^2 \nonumber \\
& & - g_l \sum_{\bm{k}\ne\bm{k'}} 
f(\bm{k}) f(\bm{k'}) c_{l,\bm{k} \uparrow}^\dagger 
c_{l, -\bm{k} \downarrow}^\dagger c_{l, -\bm{k'} \downarrow} 
c_{l,\bm{k'} \uparrow},
\end{eqnarray}
with $\xi_{\bm{k}} = -2t(\cos k_x + \cos k_y) - \mu$.
(Hereafter we set the lattice constant to unity.)
The hopping of electrons within each layer is restricted
to the nearest neighbors given by $t$.
Energies are measured in units of $t$ in the following analysis.
We focus on the half-filing case in each layer so that 
we set the chemical potential $\mu=0$.
This means that we neglect charge redistribution between the two layers. 
The second term of ${\cal H}_l$ describes the AF interaction.
The operator
$c_{lj\sigma}^\dagger$ ($c_{lj\sigma}$) creates (annihilates)
electrons at site $j$. 
The third term of ${\cal H}_l$ 
with $f(\bm{k}) = \left(\cos k_x-\cos k_y\right)/2$
describes the interaction
for $d_{x^2-y^2}$-wave pairing.

We define the $d$-wave SC order parameter in the $l$ layer with $N$ sites by
\begin{equation}
  \Delta_l = 
  \frac{1}{N} \sum_{\bm{k}} \Delta_l(\bm{k}),
  \label{eq_Delta}
\end{equation}
with 
\begin{equation}
  \Delta_l(\bm{k})=
  f(\bm{k}) \langle c_{l,\bm{k} \uparrow} c_{l, -\bm{k} \downarrow} \rangle.
\end{equation}
%We assume that the two layers have the same number of lattice sites.
The AF order parameter in the $l$ layer is defined by
\begin{equation}
  m_l = \frac{1}{2N} \sum_{\bm{k} \in {\rm RBZ}}  
  \langle
  c_{l,\bm{k}\uparrow}^{\dagger}c_{l,\bm{k}+\bm{Q}\uparrow}
  -c_{l,\bm{k}\downarrow}^{\dagger}c_{l,\bm{k}+\bm{Q}\downarrow}
  \rangle
  +\mathrm{c.c.},
  \label{eq_m}
\end{equation}
where the summation with respect to $\bm{k}$ is taken
over the reduced Brillouin zone $|k_x| + |k_y| < \pi$
and the nesting vector is $\bm{Q}=(\pi,\pi)$.
By using the order parameters, the mean field Hamiltonian 
at the $l$ layer reads
\begin{equation}
{\cal H}_l^{\rm mf} = \sum_{\bm{k}\in\text{RBZ}}C_{l \bm{k}}^\dagger
M_{l \bm{k}} C_{l \bm{k}} 
+ 4NV_l m_l^2 + Ng_l |\Delta_l|^2,
\end{equation}
where $C_{l \bm{k}} = \left( 
\begin{array}{cccc} 
  c_{l,\bm{k} \uparrow} & c_{l, \bm{k}+\bm{Q} \uparrow} & 
  c_{l, -\bm{k} \downarrow}^\dagger & c_{l, -\bm{k}-\bm{Q} \downarrow}^\dagger
\end{array} 
\right)^T$
and
\begin{equation}
M_{l \bm{k}} = 
\left(\begin{array}{cccc}
  \xi_{\bm{k}} & -4m_lV_l & - g_{\ell} \Delta_l(\bm{k}) & 0 \\
  -4m_lV_l & \xi_{\bm{k}+\bm{Q}} & 0 & - g_{\ell} \Delta_l(\bm{k}) \\
  - g_{\ell} \Delta_l(\bm{k})^\ast & 0 & -\xi_{\bm{k}} & -4m_lV_l \\
  0 & - g_{\ell} \Delta_l(\bm{k})^\ast & -4m_lV_l & -\xi_{\bm{k}+\bm{Q}}
\end{array}\right).
\end{equation}
Based on the mean field Hamiltonian of the whole system, 
$\sum_l {\cal H}_l^{\rm mf} + {\cal H}_\perp$, 
we solve the mean-field Eqs. (\ref{eq_Delta}) and (\ref{eq_m})
numerically using the $100 \times 100$ discretized Brillouin zone.

\section{Coexistence phase in single layer system}
\label{sec_single}
Before going into the analysis of the bilayer system,
we examine a coexisting phase in a single layer system
described by ${\cal H}_l^{\rm mf}$.
Reflecting the difference in symmetry 
between the AF gap created by $m_l \neq 0$ 
and the SC gap $\Delta_l$,
the coexistence phase of SC and AF orders
can be stabilized.\cite{Kyung00,Tobijaszewska2005}
The situation is similar to the slave-boson mean field
theory of the $t$-$J$ model~\cite{Inaba96} 
and the Hubbard model in the strong-coupling limit.\cite{Inui88}

\begin{figure}[tp]
  \begin{center}
    \includegraphics[width=0.8 \linewidth]{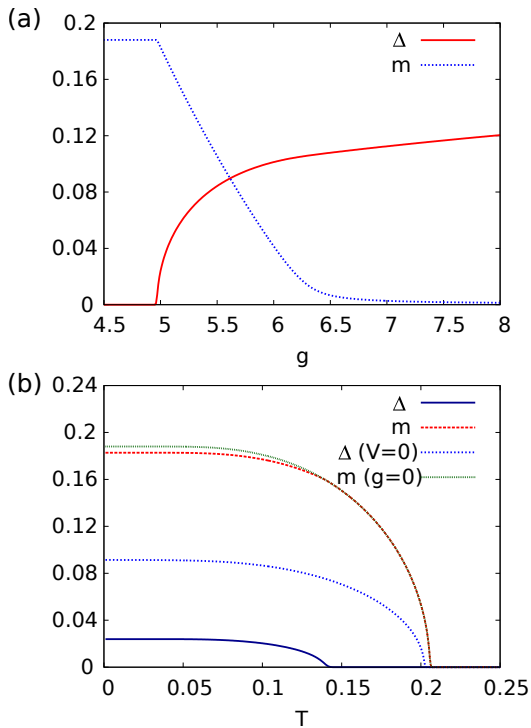}
  \end{center}
  \caption{(Color online) 
    (a) The AF order parameter $m$ and the $d$-wave SC order 
    parameter $\Delta$ 
    versus the SC interaction parameter $g$ at $V=0.5$ in the single layer
    system at $T=0$.
    (b) The temperature dependence of the order parameters
    in the coexistence phase for $g=5$ and $V=0.5$.
    For comparison we show the pure SC case ($g=5$ and $V=0$)
    and the pure AF case ($g=0$ and $V=0.5$) as well.
  }
  \label{fig_single}
\end{figure}

Figure \ref{fig_single}(a) shows the parameter range of $g\equiv g_1$
for the coexistence phase at $V\equiv V_1=0.5$.
For $g<5.0$ the system is a pure AF state
while for $g>6.3$ the system is a pure SC state.\cite{comment_Fig1a}
%{\cred
%In Fig.~\ref{fig_single}(a) $m$ has tiny values for $g>6.3$
%but this is a finite size effect.
%By increasing the number of Brillouin zone points
%taken in the numerical calculation, the finite values of $m$
%for $g>6.3$ are suppressed.
%}
So the coexistence phase appears for $5.0<g<6.3$.
For the case of $V=0.4$, the parameter for the coexistence
changes as $3.7<g<4.5$.
%{\cred
In order to confirm that the state with $\Delta \neq 0$
and $m \neq 0$ is the global minimum of the free energy,
we computed the following energy at $T=0$
for $0 \leq \Delta \leq 0.2$ and $0 \leq m \leq 0.25$:
\begin{equation}
  E=\sum_{\alpha, E_\alpha<0} E_\alpha + 4NVm^2 + Ng|\Delta|^2,
\end{equation}
where $\alpha$ runs over the all eigenstates of the mean field Hamiltonian
and $E_\alpha$ are the eigenenergies.
We examined several cases and confirmed that the coexistence phase
solution corresponds to the global minimum of the energy.
%}
We examined the $s$-wave case as well, but there is 
no coexistence phase.

There are two types of coexistence phases.
One is the phase with strong AF order and weak SC order, resulting in 
the AF transition temperature $T_{\rm AF}>T_{\rm c}$, 
and the other is the phase with $T_{\rm AF}<T_{\rm c}$.
Figure~\ref{fig_single}(b) shows the temperature 
dependence of the order parameters in the coexistence phase
with $T_{\rm AF}>T_{\rm c}$ at $g=5$ and $V=0.5$.
For comparison, the pure AF case at $g=0$ and $V=0.5$ and the pure SC case at $g=5$ and $V=0$ are also shown.
%In spite of almost the same transition temperatures in the pure cases, the AF order is selected with decreasing temperature. This is due to full-gap opening in the AF state that can gain energy as compared to the $d$-wave SC order with the node in the gap. If the SC interaction is strong, it is possible to deform the gapped dispersion toward the $d$-wave shape below $T_{\rm AF}$. As a result, the SC order appears with the reduction of $T_{\rm c}$ as compared with the pure SC case value.
For this choice of the parameters, $T_{\rm AF}$ is slightly higher than
$T_{\rm c}$.
Therefore, the system first exhibits AF order upon decreasing temperature.
$T_{\rm c}$ is somewhat reduced because of the presence of AF order.
The occurrence of the SC order also affects the AF order: the temperature dependence of $m$ in Fig.~\ref{fig_single}(b) deviates from the pure AF case at $T_{\rm c}$, resulting in the reduction of $m$. This behavior is in contrast to the case of the multilayer cuprates where the enhancement of the AF order is observed. \cite{Shimizu11}
We note that, in the coexistence phase with $T_{\rm c} > T_{\rm AF}$, $T_{\rm c}$ is the same as the value of the pure SC case but $T_{\rm AF}$ is reduced from the pure AF case value.

\section{Proximity effect in bilayer system}
\label{sec_bilayer}
In this section, we study the bilayer system.
Our purpose is twofold.
First, we study the proximity effect in the bilayer system.
Second, we examine the stability of the coexistence phase
in a single layer under the presence of inter-layer tunneling.

\begin{figure}[tp]
  %\begin{center}\includegraphics[width=0.8 \linewidth]{./images2/fig_tp.eps}
  \begin{center}\includegraphics[width=0.8 \linewidth]{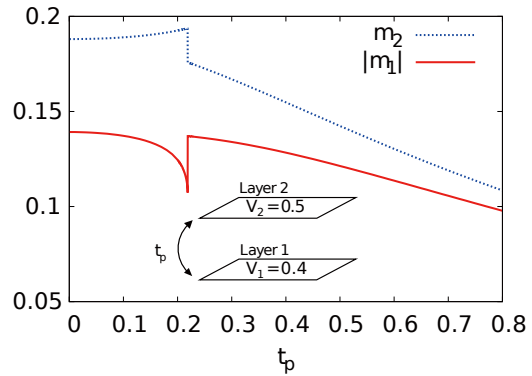}
  %\begin{center}\includegraphics[width=0.8 \linewidth]{fig2tm_ink.eps}
  \end{center}
  \caption{(Color online) 
    The AF order parameters $m_1$ and $m_2$ versus 
    the interlayer hopping parameter $t_p$ in the bilayer system for 
    $V_1=0.4$ and $V_2=0.5$.
    The inset is a schematic view of the system.
  }
  \label{fig_tp}
\end{figure}

To start with, we examine the interlayer tunneling effect.
Depending on the value of $t_p$, there are
a strong $t_p$ regime and a weak $t_p$ regime.
Figure \ref{fig_tp} shows the $t_p$ dependence 
of the AF order parameters $m_1$ and $m_2$ 
at $V_1=0.4$ and $V_2=0.5$ with $g_1=g_2=0$.
For $t_p<0.22$, we see that
the values of the order parameters are 
not so much affected by the increase of $t_p$.
This weak $t_p$ regime is not suitable for describing
multilayer systems because each layer is almost independent.
In fact, the change of $m_1$ and $m_2$ in the weak $t_p$ regime
is described by the second-order perturbation theory 
with respect to $t_p$.
At $t_p=0.22$,
there is a first order transition between
the weak $t_p$ regime and the strong $t_p$ regime 
as shown in Fig.~\ref{fig_tp}.
For $t_p>0.22$, the order parameters exhibit strong
$t_p$ dependence.
In this strong $t_p$ regime,
$t_p$ is larger than the excitation gap created by AF order.
Therefore, the order parameters are reduced due to
the change of the Fermi-surface topology.
In the large $t_p$ limit, the noninteracting single-body 
electron states are well described by the bonding state 
and the anti-bonding state.
The Fermi surface splits into two pockets centered 
at the $\Gamma$ point and $M$ point.
Qualitatively similar behaviors are found
in the $t_p$ dependence of the SC order parameters.
In the following analysis, we focus on 
this strong $t_p$ regime and set $t_p=0.3$.

Now we investigate the bilayer system.
We consider three cases.
In all cases, we assume 
$V_2 = 0.5$ 
and $g_2 = 0$ for $l=2$.
Therefore, the intrinsic order in the $l=2$ layer is restricted to AF order.
%A non-zero value of $g_2$ is necessary 
%to define the SC order parameter in the $l=2$ layer, but
%is not enough to induce intrinsic SC order.
For the $l=1$ layer, we assume $g_1 \ge 3$ and $V_1 \ge 0$.

The three cases are 
(i) $V_1=0$,
(ii) $V_1=0.4 (<V_2)$,
and (iii) $V_1=0.6 (>V_2)$.
It turns out that the ground state of
case (i) consists of SC order in the $l=1$ layer and AF order in the $l=2$ layer 
% ``sh_ms1_1g1c.sh'' and ``sh_ms1_1g1c2.sh''
% codes3/saved_data/datg1c0228_2012
when $g_1\ge 3.7$. 
The ground states of cases (ii) and (iii) are
the coexistence state of SC and AF orders in the $l=1$ layer
and only AF order in the $l=2$ layer.
Cases (ii) and (iii) are distinguished by the strength
of the AF order in the two layers:
$m_1<m_2$ in case (ii), while $m_1>m_2$ in case (iii).

\begin{figure}[tp]
  \begin{center}
    \includegraphics[width=0.7 \linewidth]{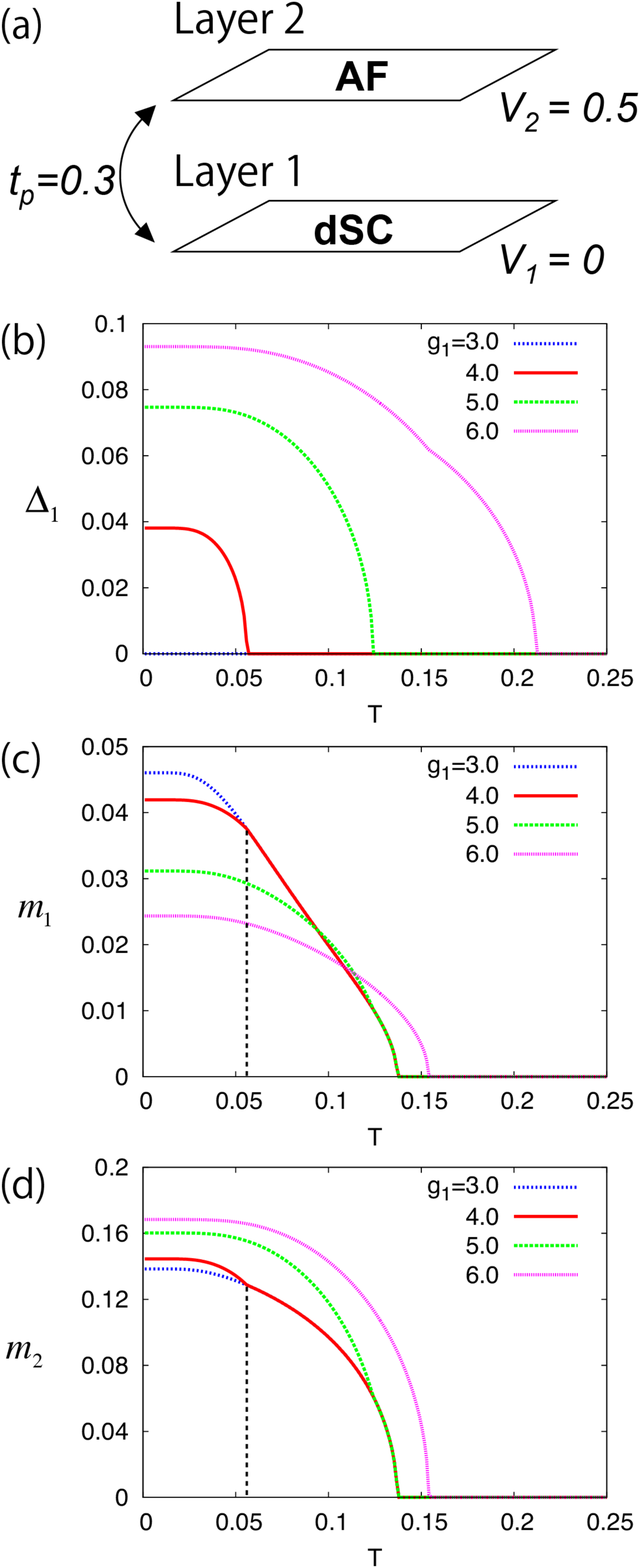}
  \end{center}
  \caption{
    (Color online) 
    (a) Schematic view of the system for the case (i) (see the text).
    The temperature dependence of the SC order parameter 
    $\Delta_1$ in layer-1 (b),
    the AF order parameter $m_1$ in layer-1 (c),
    and 
    the AF order parameter $m_2$ in layer-2 (d)
    for various $g_1$.
    The vertical dashed line represents $T_{\rm c}$ in the case of $g_1=4.0$.
  }
  \label{fig_case_i}
\end{figure}

Figure~\ref{fig_case_i}(a) is a schematic view of case (i).
The temperature dependence of the order parameters
is shown in Figs.~\ref{fig_case_i}(b)-(d)
for different values of $g_1$.
For $g_1=3$, there is no SC order in the $l=1$ layer.
% gpf_ms1_0Ta0224_2012v2.plt 
% saved_data/Data0224_2012/datNk100g3
Intrinsic AF order $m_2$ appears in the $l=2$ layer for $T<T_{\rm AF}=0.14$
as shown in Fig.~\ref{fig_case_i}(d).
As a result of the proximity effect, a finite value of $m_1$ is induced
as shown in Fig.~\ref{fig_case_i}(c).
This induced $m_1$ decreases with increasing $g_1$, because the intrinsic SC order by $g_1$ competes with the induced AF order. 
On the other hand, the intrinsic AF order $m_2$ in the $l=2$ layer
increases with increasing $g_1$.
At $g_1=4.0$ there is a SC transition in the $l=1$ layer
as shown in Fig.~\ref{fig_case_i}(b).
%{\cred
  In the presence of the non-zero SC order parameter, $\Delta_1$,
  in the $l=1$ layer,
  a finite value of $\Delta_2$ is induced
  in the $l=2$ layer (not shown) because of the proximity effect.
%}
The SC transition affects $m_2$. 
As shown in Fig.~\ref{fig_case_i}(d),
there is a clear enhancement of
$m_2$ below the SC transition temperature $T_{\rm c}=0.06$ for the case of $g_1=4$.
A similar enhancement is also found in the case of $g_1=5$
below $T_{\rm c}=0.12$.
This enhancement of $m_2$ below $T_{\rm c}$ 
is consistent with the experiment 
in the multilayer cuprate.\cite{Shimizu11}
For $g_1=6$ this enhancement is masked 
because $T_{\rm c}$ is higher than $T_{\rm AF}$.

\begin{figure}[tp]
  \begin{center}
    \includegraphics[width=0.7 \linewidth]{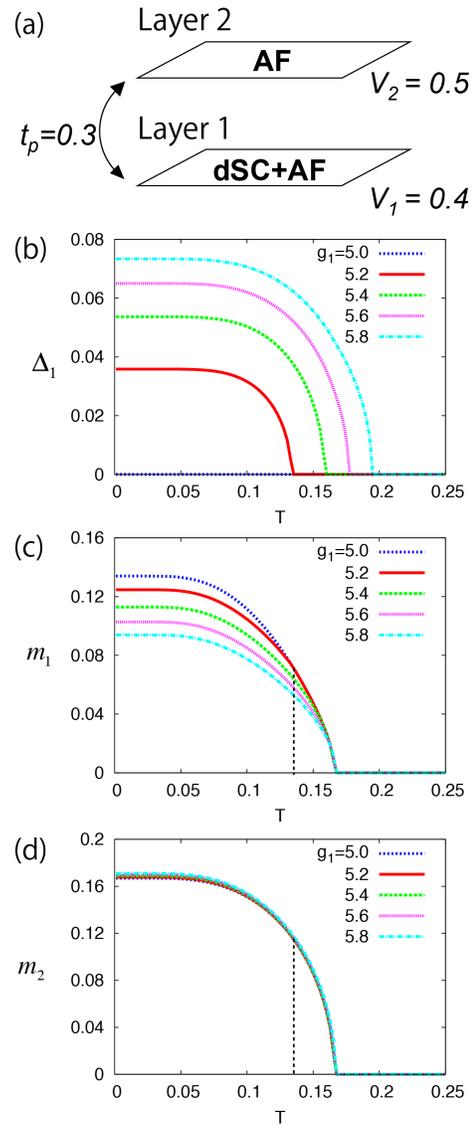}
  \end{center}
  \caption{
    (Color online) 
    (a) Schematic view of the system for the case (ii) (see the text).
    The temperature dependence of the SC order parameter 
    $\Delta_1$ in layer-1 (b),
    the AF order parameter $m_1$ in layer-1 (c),
    and 
    the AF order parameter $m_2$ in layer-2 (d)
    for different $g_1$.
    The vertical dashed line represents $T_{\rm c}$ in the case of $g_1=5.2$.
    The inset in (d) is the enlarged drawing of $m_2$ at low temperatures
    for $g_1=5.0,5.2$.
  }
  \label{fig_case_ii}
\end{figure}

The enhancement of AF order below $T_{\rm c}$
is also found in the case (ii) (Fig.~\ref{fig_case_ii}(a)).
Figures~\ref{fig_case_ii}(b)-(d) show
the temperature dependence of the order parameters
for different values of $g_1$.
For the case of
% sh_ms1_1g1cSCAF_AF.sh
% saved_data/dat0229_2012a
$g_1<5.1$,
there is no intrinsic SC order in the $l=1$ layer.
For the case of $g_1=5.2$, there is a SC transition
at 
% sh_ms1_1TbcNk100.sh 
% gpf_ms1_0TbNk100.plt
% codes3/datNk100g5_2b
$T_{\rm c}=0.13$.
Below $T_{\rm c}$, $m_2$ is clearly enhanced,
although the enhancement is much reduced compared to
the case (i).
Similar behaviors are observed for
$g_1 \geq 5.1$.
Again this behavior is consistent with the experiment.\cite{Shimizu11} 

\begin{figure}[tp]
  \begin{center}
    \includegraphics[width=0.7 \linewidth]{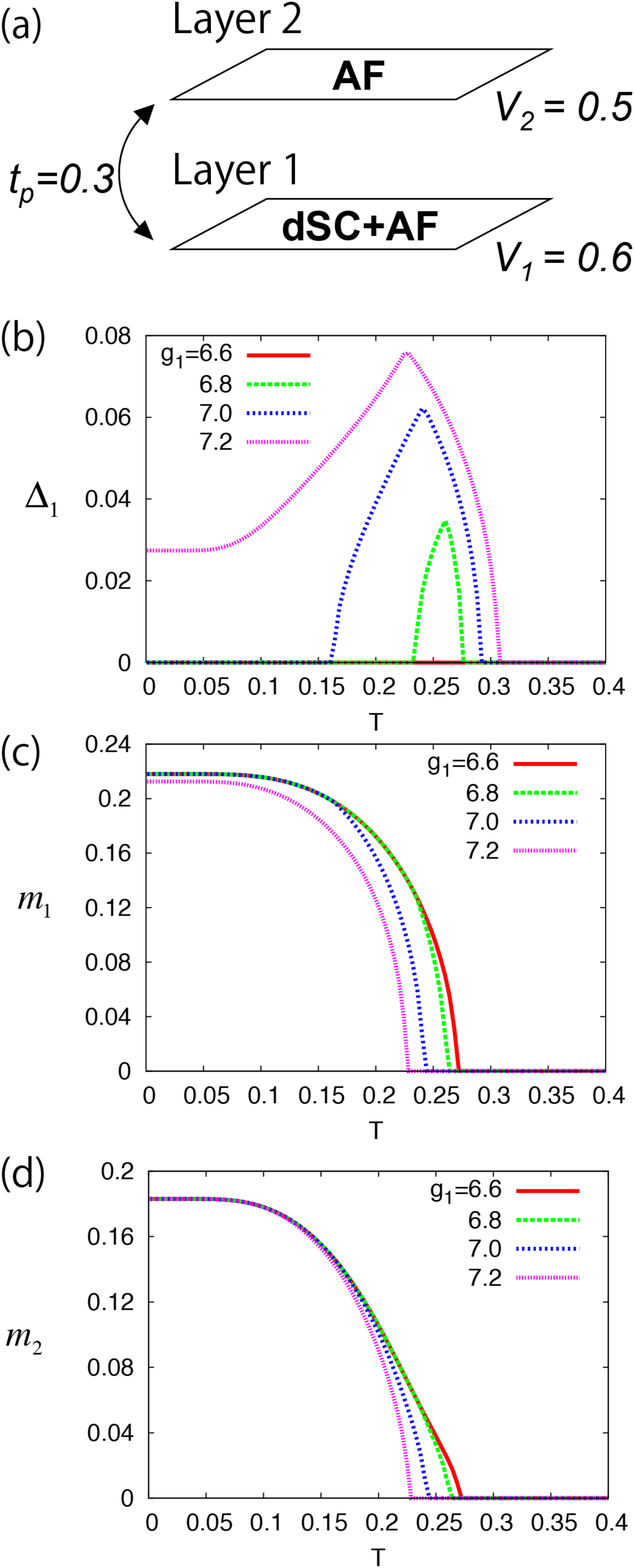}
  \end{center}
  \caption{
    (a) Schematic view of the system for the case (iii) (see the text).
    The temperature dependence of the SC order parameter 
    $\Delta_1$ in layer-1 (b),
    the AF order parameter $m_1$ in layer-1 (c),
    and 
    the AF order parameter $m_2$ in layer-2 (d)
    for different $g_1$.
  }
  \label{fig_case_iii}
\end{figure}

Now we examine the case (iii) (Fig.~\ref{fig_case_iii}(a)).
The temperature dependence of the order parameters is
shown in Figs.~\ref{fig_case_iii}(b)-(d).
In this case, we observe quite different behaviors of the order parameters
compared to the cases (i) and (ii).
In particular, the superconducting transition temperature $T_c$
is always larger than the AF transition temperature $T_{\rm AF}$.
There is no coexistence phase when $T_c < T_{\rm AF}$.
For $T_{\rm AF} < T < T_c$, $\Delta_1$ increases 
as $T$ decreases as shown in Fig.~\ref{fig_case_iii}(b).
Below $T_{\rm AF}$, $\Delta_1$ is suppressed.
Furthermore, the coexistence phase is limited to a finite range
of temperature for
% codes3/sh_ms1_1TcNk100b.sh
$6.7 \leq g_1 \leq 7.1$.  %{\cred (*** check***)} 
Meanwhile the order parameters $m_1$ and $m_2$ increase monotonically 
as the temperature decreases as shown in 
Figs.~\ref{fig_case_iii}(c) and (d).
These temperature dependences are qualitatively different from 
the experimentally observed one.
From this observation one may conclude that 
it is unlikely that there is
a coexistence phase of intrinsic SC and AF orders 
in a layer among coupled multilayers.
%{\cred
%Although there appears the coexistence phase of AF order and SC
%in all cases (i), (ii), and (iii),
Although the coexistence phase of AF and SC orders appears
in all cases (i)-(iii), 
the origin of AF order in the SC layer is different.
What makes the difference between the case (iii) and the case (ii)
is that in the case (iii) AF order in the $l=1$ layer with SC 
is intrinsic order but not induced by the other layer.
Meanwhile in the case (ii) AF order in the $l=1$ layer with SC 
is induced order by AF order in the $l=2$ layer
due to the proximity effect.
%}

\begin{figure}[tp]
  \begin{center}
    \includegraphics[width=0.7 \linewidth]{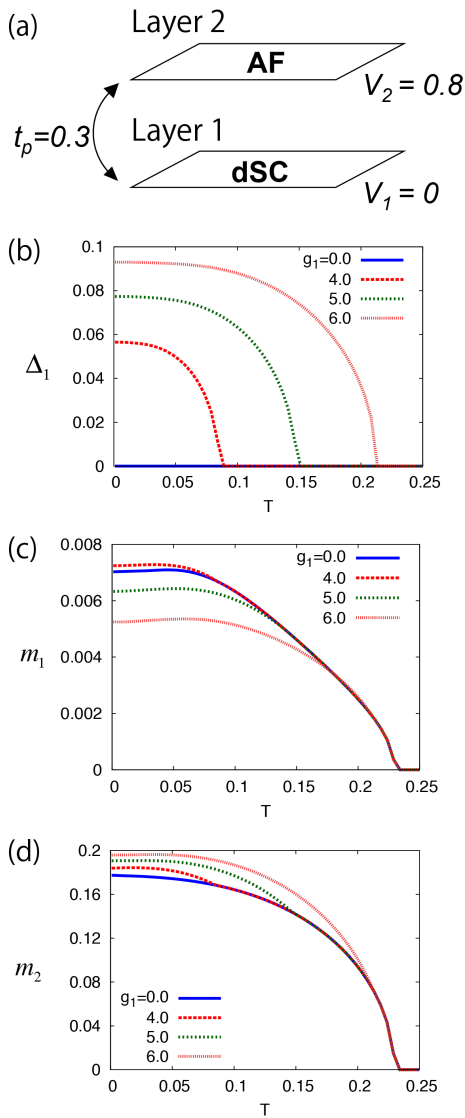}
  \end{center}
  \caption{
    (Color online) 
    (a) Schematic view of the system for the case (i) with
    $V_1=0$ and $V_2=0.8$.
    The temperature dependence of the SC order parameter 
    $\Delta_1$ in layer-1 (b),
    the AF order parameter $m_1$ in layer-1 (c),
    and 
    the AF order parameter $m_2$ in layer-2 (d)
    for various $g_1$.
  }
  \label{fig_case_i_fs}
\end{figure}

% result with realistic Fermi surface
%{\cred
So far we have studied the system with the electron hopping
restricted to the nearest neighbors.
In order to examine the system 
with a realistic Fermi surface,\cite{Damascelli2003,Kordyuk2006,Tohyama2000}
we consider the model with
\begin{eqnarray}
\xi_{\bm{k}} &=& -2t(\cos k_x + \cos k_y) 
-4t' \cos k_x  \cos k_y \nonumber \\
& & -2t'' (\cos 2k_x + \cos 2k_y) - \mu.
\end{eqnarray}
Here we choose $t'/t=-0.20$ and $t''/t=0.10$.
We take the chemical potential as $\mu=-0.839$,
which corresponds to $10\%$ doping in the normal and non-magnetic state.
The result for the case (i) above with $V_1=0$ and $V_2=0.8$
is shown in Fig.~\ref{fig_case_i_fs}.
The temperature dependence of the order parameters
is shown in Figs.~\ref{fig_case_i_fs}(b)-(d)
for different values of $g_1$.
We observe qualitatively similar behaviors of the order parameters
to the case (i) shown in Fig.~\ref{fig_case_i}.
There is a clear enhancement of $m_2$ below the SC transition.
Intrinsic AF order $m_2$ appears in the $l=2$ layer for $T<T_{\rm AF}=0.23$
as shown in Fig.~\ref{fig_case_i_fs}(d).
This intrinsic order is enhanced below the SC transition.
For example, there is the SC transition at $T=0.15$ for $g_1=5.0$
as shown in Fig.~\ref{fig_case_i_fs}(b).
For $T<0.15$, $m_2$ is enhanced compared with the $g_1=0$ case.
Meanwhile, $m_1$ is suppressed as shown in Fig.~\ref{fig_case_i_fs}(c).
The exceptional case is $g_1=4.0$.
The temperature dependence of $m_2$ is similar to the other cases but 
the temperature dependence of $m_1$ is qualitatively different.
The value of $m_1$ is enhanced below the SC transition.
This discrepancy is probably associated with the change of 
the Fermi surface shape.
%However, this behavior is observed only for $2.5 < g_1 < 4.5$.
%For $g_1 > 4.5$, we observe qualitatively the same behavior
%as that shown in Fig.~\ref{fig_case_i}.
%}

\section{Summary}
\label{sec_summary}
To summarize, we have studied the proximity effect and the possibility
of coexistence of AF and SC orders in a bilayer system.
Our mean field theory suggests that
the experimentally observed enhancement of AF order
below $T_{\rm c}$~\cite{Shimizu11} 
is associated with the proximity effect.
In contrast, if we assume a coexistence phase
in a layer among coupled multilayers,
the temperature dependence of the order parameters
is qualitatively different from the experimentally observed one.

%{\cred
We believe that this result is not so much affected by
the shape of the Fermi surface.
As we have shown in Fig.~\ref{fig_case_i_fs},
the result for a realistic Fermi surface with finite doping
is qualitatively the same as that for the half-filling case.
%}
So we expect qualitatively the same proximity effect as long as 
we neglect the possibility of stabilizing other orders,
such as a charge-density wave or 
the so-called $\pi$-triplet pairing.\cite{Psaltakis83,Murakami98,Kyung00}
The absence of the $\pi$-triplet pairing is the unique property
of the half-filling case with $t'=t''=0$.
However, there is no experimental 
evidence for the $\pi$-triplet pairing
to the best of our knowledge.

\section*{Acknowledgment}
This work was supported by the Grant-in-Aid for Scientific Research 
from the Ministry of Education, Culture, Sports, Science,
and Technology of Japan; the Global Center Of Excellence (COE) Program 
``The Next Generation of Physics, Spun from University and Emergence";
and the Yukawa Institutional Program for Quark-Hadron Science 
at Yukawa Institute of Theoretical Physics.


\begin{thebibliography}{99}
%%% multilayer systems
\bibitem{Ihara88}
  H.~Ihara, R.~Sugise, M.~Hirabayashi, N.~Terada, M.~Jo, K.~Hayashi, A.~Negishi, M.~Tokumoto, Y.~Kumira, and T.~Shimomura, Nature (London) {\bf 334}, 510 (1988).

\bibitem{Parkin88}
S.~S.~P.~Parkin, V.~Y.~Lee, A.~I.~Nazzal, R.~Savoy, and R.~Beyers, S.~J.~La~Placa Phys. Rev. Lett. {\bf 61}, 750 (1988).

\bibitem{Tokunaga99} 
Y. Tokunaga, H. Kotegawa, K. Ishida, G. -q. Zheng, Y. Kitaoka, K. Tokiwa, A. Iyo and H. Ihara, 
J. Low Temp. Phys. \textbf{117}, 473 (1999).

\bibitem{Kotegawa04}
H. Kotegawa, Y. Tokunaga, Y. Araki, G. -q. Zheng, Y. Kitaoka, K. Tokiwa, K. Ito, T. Watanabe, A. Iyo, Y. Tanaka and H. Ihara, 
Phys. Rev. B \textbf{69}, 014501 (2004). 

\bibitem{Mukuda06} 
H. Mukuda, M. Abe, Y. Araki, Y. Kitaoka, K. Tokiwa, T. Watanabe, A. Iyo, H. Kito and Y. Tanaka, 
Phys. Rev. Lett. \textbf{96}, 087001 (2006).

\bibitem{Mukuda08} 
H. Mukuda, Y. Yamaguchi, S. Shimizu, Y. Kitaoka, P. Shirage, and A. Iyo, 
J. Phys. Soc. Jpn. \textbf{77}, 124706 (2008).

\bibitem{Shimizu09}
S. Shimizu, H. Mukuda, Y. Kitaoka, H. Kito, Y. Kodama. P. M. Shirage and A. Iyo, J. Phys. Soc. Jpn. \textbf{78}, 064705 (2009). 

\bibitem{Inui88}
M. Inui, S. Doniach, P. J. Hirschfeld and A. E. Ruckenstein, Phys. Rev. B \textbf{37}, 2320 (1988). 

\bibitem{comment_Fig1a}
In Fig.~\ref{fig_single}(a) $m$ has tiny values for $g>6.3$
but this is a finite size effect.
By increasing the number of Brillouin zone points
taken in the numerical calculation, the finite values of $m$
for $g>6.3$ are suppressed.

\bibitem{Chen90}
G. J. Chen, R. Joynt, F. C. Zhang and C. Gros, Phys. Rev. B \textbf{42}, 2662 (1990). 

\bibitem{Giamarchi91}
T. Giamarchi and C. Lhuillier, Phys. Rev. B \textbf{43}, 12943 (1991). 

\bibitem{Shimizu11} 
S. Shimizu, S. Tabana, H. Mukuda, Y. Kitaoka, P. M. Shirage, H. Kito and A. Iyo, 
J. Phys. Soc. Jpn. \textbf{80}, 043706 (2011). 

\bibitem{Fernandes10} 
R.~M.~Fernandes, D.~K.~Pratt, W.~Tian, J.~Zarestky, A.~Kreyssig, S.~Nandi, M.~G.~Kim, A.~Thaler, N.~Ni, P.~C.~Canfield, R.~J.~McQueeney, J.~Schmalian, and A.~I.~Goldman,
Phys. Rev. B \textbf{81}, 140501(R) (2010). 

\bibitem{Kyung00}
B. Kyung, Phys. Rev. B \textbf{62}, 9083 (2000).

\bibitem{Tobijaszewska2005}
B.~Tobijaszewska and R.~Micnas, phys. stat. sol. (b) \textbf{242},
468 (2005).

\bibitem{Inaba96}
M. Inaba, H. Matsukawa, M. Saitoh and H. Fukuyama, Physica C \textbf{257}, 299 (1996). 

\bibitem{Damascelli2003} 
A.~Damascelli, Z.~Hussain, and Z.-X.~Shen, Rev. Mod. Phys. \textbf{75}, 
473 (2003).
\bibitem{Kordyuk2006}
A.~A.~Kordyuk and S.~V.~Borisenko, Low Temp. Phys. \textbf{32}, 298 (2006).
\bibitem{Tohyama2000}
T.~Tohyama and S.~Maekawa, Supercond. Sci. Technol. \textbf{13},
R17 (2000).

\bibitem{Psaltakis83}
  G. C. Psaltakis and E. W. Fenton, J. Phys. C {\bf 16}, 3913 (1983).

\bibitem{Murakami98} 
M. Murakami and H. Fukuyama, J. Phys. Soc. Jpn. \textbf{67}, 2784 (1998). 


%*****************************************************************

\end{thebibliography}
\end{document}